\def\zem{$z_{\rm em}$}
\def\zabs{$z_{\rm abs}$}

\def\l{$\lambda$}

\def\e{et~al.}

\def\hi{H~{\sc i}}

\def\nhi{\mbox{$\sc N(\sc H~{\sc i})$}}
\def\lognhi{\mbox{$\log \sc N(\sc H~{\sc i})$}}

\def\caii{Ca~{\sc ii}}
\def\caiii{Ca~{\sc iii}}

\def\feii{Fe~{\sc ii}}

\def\mgi{Mg~{\sc i}}
\def\mgii{Mg~{\sc ii}}

\def\mnii{Mn~{\sc ii}}

\def\siii{Si~{\sc ii}}

\def\tiii{Ti~{\sc ii}}

\def\aliii{Al~{\sc iii}}
\def\znii{Zn~{\sc ii}}
\def\crii{Cr~{\sc ii}}
\def\coii{Co~{\sc ii}}

\def\nai{Na~{\sc i}}

\documentstyle[psfig,rotating]{mn}

\input{psfig.sty}

\title[Metal abundances at z$<$1.5]{Metal abundances at z$<$1.5: new measurements in sub-Damped Lyman-$\alpha$ Absorbers \thanks{Based on observations collected during programme ESO
 78.A-0646 at the European Southern Observatory with UVES on the 8.2 m
 KUEYEN telescope operated at the Paranal Observatory, Chile.} }
\author[C. P\'eroux et al.] {C. P\'eroux$^1$\thanks{e-mail:
        celine.peroux@oamp.fr}, J. D. Meiring$^2$, V. P. Kulkarni$^2$,
P. Khare$^3$, J. T. Lauroesch$^4$,
\newauthor
 G. Vladilo$^5$,  \& D. G. York$^6$.\\
$^1$ Observatoire Astronomique de Marseille-Provence, Laboratoire dÕAstrophysique de Marseille,\\ UMR6110, CNRS/Universit\'e de Provence, Marseille, France. \\
$^2$ Dept. of Physics and Astronomy, Univ. of South Carolina, Columbia, USA.\\
$^3$ Dept. of Physics, Utkal University, Bhubaneswar, India.\\
$^4$ Dept of Physics and Astronomy, Univ. of Louisville, USA.\\
$^5$ Osservatorio di Trieste, Trieste, Italy.\\
$^6$ Dept. of Astronomy and Astrophysics \& Enrico Fermi Institute, Univ. of Chicago, Chicago, USA.
}
\begin{document}

\date{Accepted 2008 March 04. Received 2008 March 03; in original form 2008 January 11}

\pagerange{\pageref{firstpage}--\pageref{lastpage}} \pubyear{2002}

\maketitle

\label{firstpage}

\begin{abstract}
Damped Lyman-$\alpha$ systems (DLAs) and sub-DLAs seen toward background
quasars provide the most detailed probes of elemental abundances.  
Somewhat paradoxically these measurements are more difficult at lower
redshifts due to the atmospheric cut-off, and so a few years ago our group
began a programme to study abundances at z $<$ 1.5 in quasar absorbers.
In this paper, we present new UVES observations of six additional quasar absorption line systems at z $<$ 1.5, five of which are sub-DLAs. 
We find {\it solar or above solar} metallicity, as measured by the abundance of zinc, assumed not to be affected by dust, in two sub-DLAs: one, towards Q0138$-$0005 with [Zn/H]=$+$0.28$\pm$0.16; the other towards Q2335$+$1501 with [Zn/H]=$+$0.07$\pm$0.34. Relatively high metallicity was observed in another system: Q0123$-$0058 with [Zn/H]=$-$0.45$\pm$0.20. Only for the one DLA in our sample, in Q0449$-$1645, do we find a low metallicity, [Zn/H]=$-$0.96$\pm$0.08. We also note that in some of these systems large relative abundance variations from component to component are observed in Si, Mn, Cr and Zn. 
\end{abstract}


\section{Introduction}

Damped Lyman-$\alpha$ systems (hereafter DLAs) seen
in absorption in the spectra of background quasars, select galaxies over
all redshifts independent of their intrinsic luminosity. They have
hydrogen column densities, $\lognhi \ga$  20.3. DLAs are
also the major contributors to the neutral gas density, $\Omega_{\rm HI}$, in
the Universe at high redshifts (P\'eroux et al. 2003a; Prochaska et al. 2005) but slightly lower column density systems are also believed to contribute to the 
\hi\ mass (P\'eroux et al. 2005). These latter systems are referred to in the community as sub-Damped Lyman-$\alpha$ systems (sub-DLAs) and were defined by P\'eroux et al. (2003a) to have 19.0 $\la \lognhi \la$  20.3. Furthermore, both DLAs and sub-DLAs offer a direct probe of element abundances over $ > 90 \%$ of the age of the Universe. The gas phase abundances, measured spectroscopically, are often depleted relative to the true abundances because some of the atoms are incorporated into dust grains. For this reason the abundances of metal-rich/dusty absorbers can only be derived from volatile elements, which show little affinity with the dust, such as sulphur and zinc. The only volatile,  nearly undepleted element accessible at low redshifts with ground-based instruments and with 
usually unsaturated lines that often lie outside the Lyman-$\alpha$ forest is zinc (Pettini et al. 1998). Indeed, lines of other elements that are normally not highly depleted in the interstellar medium, in particular oxygen, nitrogen and sulphur, but they are often saturated and are not accessible at the low redshifts we study here, except from space.

Most cosmic chemical evolution models (e.g., Malaney, \& Chaboyer 1996; Pei, Fall, \& Hauser 1999) predict the global interstellar metallicity to rise from almost 1/1000 solar at
high redshift to nearly solar at $z=0$. The present-day mass-weighted
mean interstellar metallicity of local galaxies is, in fact, nearly solar (e.g.,
Kulkarni \& Fall 2002). DLAs and sub-DLAs are expected
to follow this behaviour if they trace an unbiased sample of galaxies
selected only by \nhi. There has been considerable debate
about whether or not high \nhi\ quasar absorbers actually show this predicted rise up to solar
metallicity. This ambiguity is mainly due to the small number of
measurements available for study at $z < 1.5$, which corresponds to $\sim 70 \%$
of the age of the Universe. Measurements of the metallicities of
galaxies detected in emission in this so-called ``redshift desert''
are extremely sparse, although this redshift interval covers a
significant look-back time. Therefore, information gained in
absorption is of utmost importance at these epochs. The problem at z $<$ 1.5 is that (a) the \znii\ lines lie 
in the UV for z $<$ 0.6 and in the blue at 0.6 $<$ z $<$ 1.5, and (b) the \hi\
Lyman-$\alpha$ lines lie in the UV for z $<$ 1.5. Even at these low  redshifts, DLAs appear to have low
 metallicities (e.g., Kulkarni et al. 2005; Meiring et al. 2006; P\'eroux et al. 2006b).

Furthermore, a missing metals problem is known to exist at high
redshifts (e.g., Bouch\'e, Lehnert, P\'eroux 2005;
Bouch\'e, Lehnert, P\'eroux 2006; Pettini 2006). Indeed, no more than $\approx 30-40$\% of the
metals expected to be produced by $z = 2.5$ are actually observed if
one sums up the metals in the DLAs (which are less common and only
$\sim 1/10$ solar), the Lyman-$\alpha$ forest lines (which are much more
common but only $\sim 1/1000$ solar), and the Lyman-break galaxies
(which are relatively rare). Where are the missing metals? Some have suggested that warm/hot gas in galactic halos represents a plausible contribution to the missing metals problem (Ferrara et al. 2005; Fox et al. 2007; Sommer-Larsen et al. 2008). Another
possibility is that the more metal-rich DLAs are dustier, obscure the
background quasars more, and are therefore under-represented in the
present samples. The results of the search for DLAs towards
radio-selected quasars suggest that dust obscuration might be modest
(Ellison et al. 2001a), but the sample is still small
to derive firm conclusions.  Dust effects can be important if the
extinction increases with metal column density as found in
interstellar clouds (Vladilo et al. 2006; Vladilo \& P\'eroux 2005). Indeed, the dust
content increases with metallicity (Vladilo 2004) and dust is expected to play a significant 
role at z $<$ 1.5 if the metallicity increases with cosmic time. A way to find
metal-rich absorbers minimising dust effects is to search for quasar
absorbers with relatively low \hi\ column density. In fact, recent data
are suggesting that metal-rich systems can be found more easily in
these absorbers than in classical DLAs (e.g., Pettini et al. 2000; Jenkins et al. 2005). In particular, 
the growing sample of observations of sub-DLAs includes some high metal abundances (e.g. Khare et al.  2004; P\'eroux et al. 2006a; Meiring et al. 2007, 2008). 

In this paper, we present high resolution observations of six quasars
taken with UVES at VLT. These observations increase the sample of low-redshift sub-DLAs observed at high-resolution to determine metallicity using dust-free estimates based on zinc. The paper is structured
as follows. Target selection strategy and details of observations are
provided in Section 2. 
In the third section, we detail the analysis and the profile fit of each of our quasar absorbers, including the total abundances with respect to solar. Finally, the last section analyses these results in terms of metallicity and dust content.

\section{Data Reduction}

\begin{table*}
\begin{center}
\caption{Journal of observations. \nhi\ column
densities are in cm$^{-2}$.
\label{t:JoO}}
\begin{tabular}{ccccccccc}
\hline
\hline
 Quasar & SDSS ID &g mag$^a$ & \zem$^b$ & \zabs$^c$ & \lognhi &Obs Date & $\lambda_{\rm c}$ (nm) &T$_{\rm exp}$ (sec)\\
 \hline
Q0123$-$0058  &SDSS J012303.22$-$005818.9&18.7 &1.549 &1.4094  &20.08$^{+0.10}_{-0.08}$ &01 Oct 2006   &520	 	&2$\times$2700\\
Q0132$+$0116 &SDSS J013233.90$+$011607.2&18.8 &1.786 &1.2712 &19.70$^{+0.08}_{-0.10}$ &03 Oct 2006   &520	 	&3$\times$4800$+$1$\times$1166\\
Q0138$-$0005  &SDSS J013825.53$-$000534.5&18.8 &1.341 &0.7821  &19.81$^{+0.06}_{-0.11}$ &12/13/19 Oct 2006  &346$+$564	&3$\times$5200\\
Q0153$+$0009 &SDSS J015318.19$+$000911.3&17.9 &0.838 &0.7714 &19.70$^{+0.08}_{-0.10}$ &04 Oct 2006   &346$+$564	&2$\times$5400\\
Q0449$-$1645$^d$  &--						      &18.0 &2.679 &1.0072  &20.98$^{+0.06}_{-0.07}$ &04/14/20 Oct 2006&390$+$564	&3$\times$5200\\
Q2335$+$1501 &SDSS J233544.19$+$150118.3&18.2 &0.790 &0.6798   &19.70$^{+0.30}_{-0.30}$	             &03/11/13 Oct 2006   &346$+$564	&4$\times$4480\\
\hline
\hline
\end{tabular}
\end{center}
\vspace{0.2cm}
\begin{minipage}{140mm}
{\bf $^a$:} Magnitudes are SDSS g band when available and V band otherwise.\\
{\bf $^b$:} \zem\ is from the SDSS when available.\\
{\bf $^c$:} \zabs\ is from Rao, Turnshek \& Nestor (2006).\\
{\bf $^d$:} Q0449$-$1645 was not observed by SDSS.\\
\end{minipage}
\end{table*}

The targets for our observing programme were generally selected from the Sloan Digital Sky Survey (SDSS) sample of quasars, augmented with few objects from the literature, with \mgii\ absorbers known to have DLAs or sub-DLAs at $0.6 < z < 1.5$ with  \nhi\ measurement from Hubble Space Telescope (HST).

The data were reduced using the most recent version of the UVES
pipeline in {\tt MIDAS} (uves/2.1.0 flmidas/1.1.0). Master bias and
flat images were constructed using calibration frames taken closest in
time to the science frames. The science frames were extracted with the
``optimal'' option. The blue part of the spectra were checked order by order to verify that 
all were properly extracted. The resulting spectra were combined, weighting
each spectrum by signal-to-noise ratio, corrected to vacuum
heliocentric reference and normalised.  To perform the
abundance studies, the spectra were divided into 50-75 {\AA} regions,
and each region was normalised using cubic spline functions of orders
1 to 5 as the local continuum. The final spectra have a resolution 
R=\l/$\Delta$\l$\sim$45000 and are rebinned to $\sim$0.05 \AA/pix to obtain 
a significantly improved signal-to-noise ratio. These rebinned spectra were used for all measurements: 
Voigt profile fits, Apparent Optical Depth (AOD) measures and equivalent width measurements.

One of the objects in our programme, Q0217$+$0144,
yielded a spectrum with too low signal-to-noise ratio to be
useful and therefore is not included in our sample. This object
appears to be variable, Bergeron and D'Odorico (1986)
measured 13.8 AB magnitude at 6000 \AA\ in
Aug. 1984, Blades et al. (1985) B$>$18.5, Pettini et al. (1983)
report B=14.5-16.5 and Bolton and Wall (1969) B$>$19.5.
While at the time of submission of the observing proposal
our best estimate for the magnitude was g$\sim$16,
it dimmed by at least 3 magnitudes by the time the
observations took place. This strong variation in magnitude
suggests that the object might be a BL Lac.

\section{Analysis}

\subsection{Methodology}

In the following, each sub-section title includes the name of the quasar, the value of the \hi\ column density, the absorber redshift and the number of components. The error estimates of the \nhi\ measurements are provided in Table~\ref{t:JoO}. They are derived from Space Telescope Imaging Spectrograph (STIS) observations by Rao, Turnshek \& Nestor (2006) except for Q2335$+$1501 where Advanced Camera for Surveys (ACS) prism spectra (prog ID 10556; PI: D. Turnshek; S. Rao et al., private communication) were used.  As shown in Meiring et al. (2008), the single component fits used are degenerate in \lognhi\ in relatively low resolution, low signal-to-noise HST data, leading to artificially high \hi\ column densities. This makes our measured abundances, [X/H], conservative in the sense that they are slightly lower than the actual ones. The absorber redshifts reported in Table~\ref{t:JoO} are from Rao, Turnshek \& Nestor (2006).

The metal column densities were estimated as in Meiring et al. (2006, 2007, 2008) and P\'eroux et al. (2006a, 2006b) by fitting mutliple-component Voigt
profiles to the observed absorption lines using the program FITS6P
(Welty \e\ 1991) which evolved from the code used by
Vidal-Madjar et al. (1977). FITS6P minimises the $\chi^{2}$ between the data and the theoretical Voigt profiles convolved with the
instrumental profile. The same fitting procedure was used for all the
systems. The number of components were kept to a minimum but future higher-resolution observations might require more components than fitted in this work as indicated by relatively large $b$ values. Basically, the velocity and Doppler $b$ parameters of the various components were estimated from the \mgi, \mgii\ and \feii\ lines
when available and then kept fixed for the remaining lines, allowing for variations
from metal species to metal species in $N$ only. This procedure is standard practice, and is appropriate because the b values are not expected to show systematic trends with atomic mass 
because they are dominated by bulk motions rather than thermal broadening. The resulting error estimates for the velocities are $\sim$ 0.5--1.0 km/s and less than 1 km/s for the $b$ values. All the transitions observed were used together to get the column density for each species. In particular, the very weak \feii\ \l\ 2249, 2260 lines, even when not detected, constrain the values of N(\feii). In all cases, the \mgi\ 
$\lambda$ 2026.5 contribution to the \znii\ $\lambda$ 2026.1 line was
estimated using the component parameters for \mgi\ derived from the \mgi\
$\lambda$ 2852 profile and using $f_{\rm osc}$(\mgi\ \l\ 2852)/$f_{\rm osc}$(\mgi\ \l\ 2026)=32. In Figures~\ref{f:Q0123} to \ref{f:Q2335} for \znii\ \l\ 2026.14, the nearby \mgi\ \l\ 2026.48 line is noted by vertical dashed lines. Likewise for \crii\ \l\ 2062.23 in the plots of \znii\ \l\ 2062.66. In all cases, both pairs of ions are fitted simultaneously, thus including the contribution of the weaker line that can be estimated given the high-resolution of the data. The
atomic data were adopted from Morton (2003). All wavelengths reported here are truncated vacuum values.

In Tables~\ref{t:Q0123} to \ref{t:Q2335}, we give the resulting column densities. The few weak components, in
which the column densities could not be well constrained due to higher
noise are marked with `--'. Their contributions to the total column densities are negligible and
therefore not included in the total sum. Lower limits in a given saturated component of the profile fit were estimated from the fit. In case the components with the strongest lines of other species are noisy, we derive upper limits to the total column densities. Upper limits to the total column densities deduced from non detection were derived at 3-$\sigma$ using the local signal-to-noise estimate.

\subsection{Individual Systems}

\subsubsection{Q0123$-$0058; \lognhi=20.08; \zabs=1.4094; 10 components }

\begin{figure*}
\begin{center}
\includegraphics[height=13cm, width=13cm, angle=0]{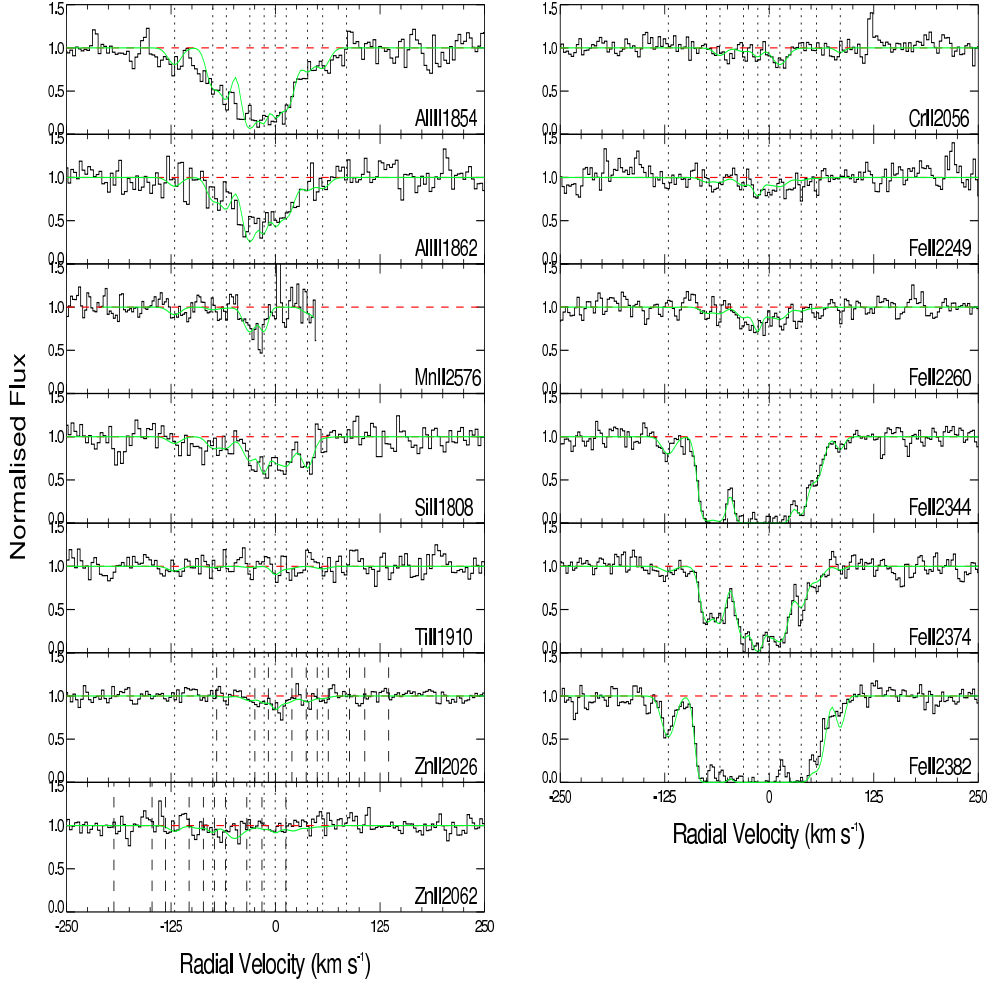}
\caption{Voigt profile fit of the \zabs=1.4094 sub-DLA towards Q0123$-$0058 
(see Table~\ref{t:Q0123}). The dashed vertical lines in the \znii\
$\lambda$ 2026 and 2062 panels indicate the position of the MgI $\lambda$
2026 and \crii\ $\lambda$ 2062 components respectively. The spectra are rebinned to 0.05\AA/pix.}
\label{f:Q0123}
\end{center}
\end{figure*}

\begin{table*}
\begin{center}
\caption{Parameters for the Voigt profile fit of the $z_{\rm abs} = 1.4094$ sub-DLA towards 
Q0123$-$0058 (see Figure~\ref{f:Q0123}). Components marked "--" are not detected in 
the corresponding transition. Velocities and $b$ values are in km s$^{-1}$ and column
densities are in cm$^{-2}$.}
\label{t:Q0123}
\begin{tabular}{cccccccc}
\hline\hline
         Vel      & b    &N(\aliii)   &N(\siii)   &N(\mnii)  &N(\crii)   &N(\feii)   &N(\znii) \\  
\hline
$-$120.7	&	9.9	&	(1.49$\pm$0.34)E12	&	(1.55$\pm$1.23)E14	&	--			&	--			&	(5.84$\pm$0.63)E12	&	--	\\
$-$75.0	&	8.7	&	(3.73$\pm$0.48)E12	&	(2.41$\pm$1.29)E14	&	--			&	--			&	(8.25$\pm$0.53)E13	&	--	\\
$-$58.9	&	8.6	&	(5.40$\pm$0.57)E12	&	(2.40$\pm$1.28)E14	&	--			&	--			&	(8.51$\pm$0.55)E13	&	--	\\
$-$30.7	&	9.8	&	(1.89$\pm$0.17)E13	&	(6.20$\pm$1.57)E14	&	(2.50$\pm$0.87)E12	&	--			&	(1.66$\pm$0.09)E14	&	--	\\
$-$13.7	&	6.2	&	(9.15$\pm$1.28)E12	&	(7.04$\pm$1.70)E14	&	(1.56$\pm$0.72)E12	&	--			&	(2.76$\pm$0.45)E14	&	(3.82$\pm$1.84)E11	\\
$-$0.5	&	5.7	&	(5.56$\pm$0.98)E12	&	(2.63$\pm$1.48)E14	&	--			&	--			&	(5.62$\pm$0.95)E13	&	(6.32$\pm$2.03)E11	\\
12.8		&	11.5	&	(9.85$\pm$0.88)E12	&	(9.04$\pm$1.95)E14	&	--			&	(8.18$\pm$1.83)E12	&	(2.20$\pm$0.12)E14	&	(6.90$\pm$2.49)E11	\\
38.2		&	8.3	&	(1.78$\pm$0.37)E12	&	(7.47$\pm$1.64)E14	&	--			&	--			&	(4.96$\pm$0.25)E13	&	-	\\
56.4		&	9.9	&	(1.86$\pm$0.37)E12	&	--			&	--			&	--			&	(1.81$\pm$0.10)E13	&	--	\\
85.0	&	6.2	&	--			&	--			&	--			&	--			&	(2.80$\pm$0.50)E12	&	--	\\
\hline 				       			 	 
\hline 				       			 	 
\end{tabular}			       			 	 
\end{center}			       			 	 
\end{table*}			       			 	 

A number of features are detected over some or all of the profile components: \aliii\ \l\ 1854, 1862; \siii\ \l\ 1808; \mnii\ \l\ 2576; \crii\ \l\ 2056; \feii\ (\l \l\ 2249, 2260, 2344, 2374, 2382) and \znii\ \l \l\ 2026, 2062. \tiii\ \l\ 1910 is covered but not detected. Based on the SDSS DR3 spectra, we note that the \aliii\ \l\ 1854 is blended with the \mnii\ \l\ 2594 line from a lower redshift strong \mgii\ absorber (\zabs=0.723). In this system, it is interesting to note that the column density ratios of Si, Mn, Cr and Zn vary from one component to another. By comparison, the ratio of column densities of \aliii\ and \feii\ is about 1.2 in all components, close to the mean found by Meiring et al. (2007, 2008). Assuming no variations in the radiation field, relative abundances of the noted elements seem to be variable from component to component. The results of the profile fitting analysis are summarised in Table~\ref{t:Q0123}, while Figure~\ref{f:Q0123} illustrates the fits.

\subsubsection{Q0132$+$0116; \lognhi=19.70; \zabs=1.2712; 11 components }  

\begin{figure*}
\begin{center}
\includegraphics[height=13cm, width=13cm, angle=0]{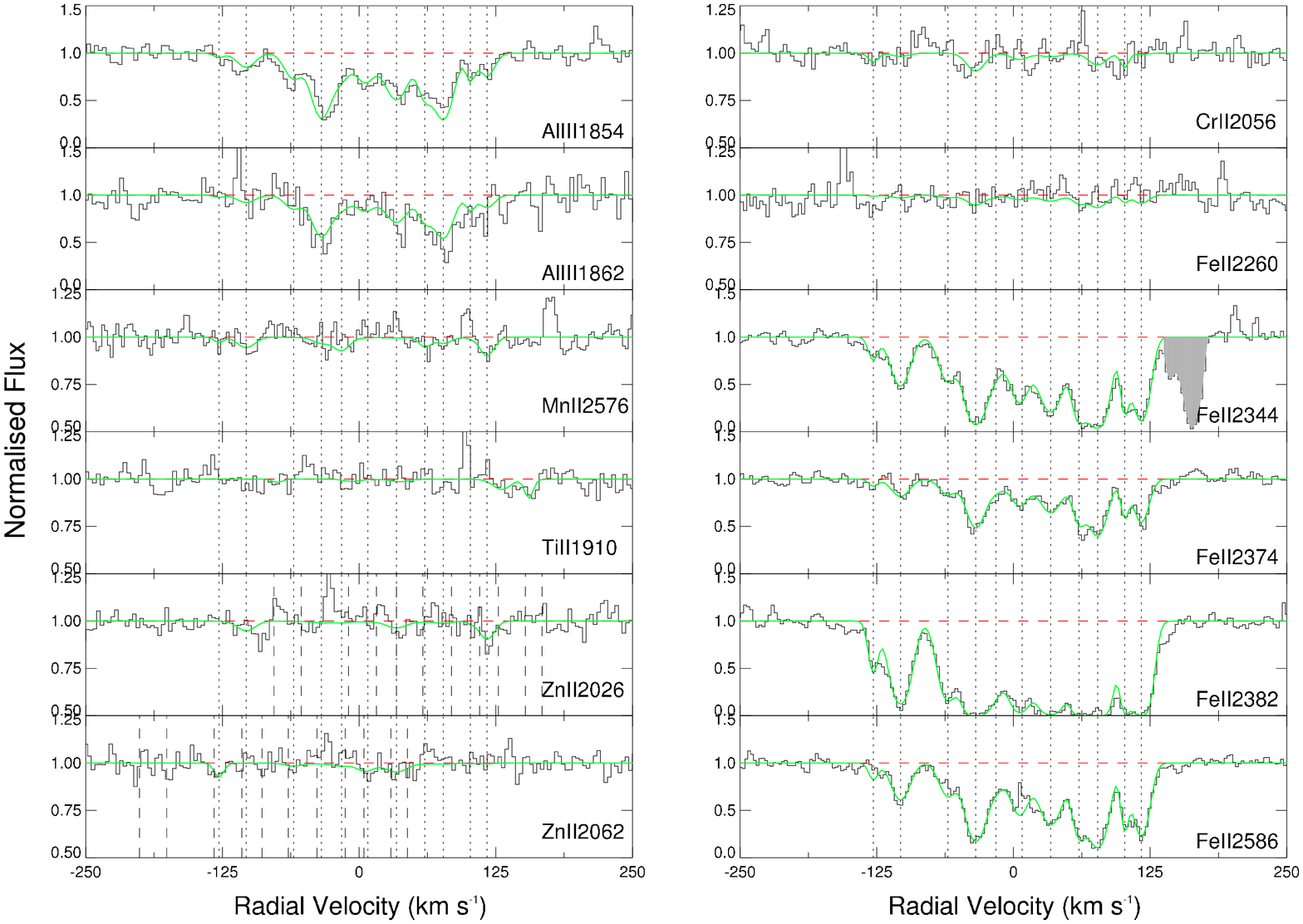}
\caption{Voigt profile fit of the \zabs=1.2712 sub-DLA towards Q0132$+$0116 (see Table~\ref{t:Q0132}). \feii\ \l\ 2344 is blended with the \mgii\ \l\ 2796 line from a known \mgii\ system \zabs=0.9050 (shaded in grey). The peaks at v=$+$100 km s$^{\rm -1}$ in \tiii\ \l\ 1910 and \mnii\ \l\ 2576 and v=$+$190 km s$^{\rm -1}$ in \mnii\ \l\ 2576 are due to bad sky subtraction in the reduction process. \label{f:Q0132}}
\end{center}
\end{figure*}

\begin{table*}
\begin{center}
\caption{Parameters for the Voigt profile fit of the $z_{\rm abs} = 1.2712$ sub-DLA towards Q0132$+$0116 (see Figure~\ref{f:Q0132}). }
\label{t:Q0132}
\begin{tabular}{cccccccc}
\hline\hline
         Vel      & b    &N(\aliii)   &N(\feii)   \\ 
\hline
$-$128.1	&4.9	&-- 	    	    &(4.20$\pm$0.71)E12\\
$-$103.2	&10.7	&(1.21$\pm$0.31)E12 &(2.07$\pm$0.13)E13\\ 
$-$59.9		&10.1	&(2.27$\pm$0.36)E12 &(1.71$\pm$0.13)E13\\ 
$-$34.5		&10.5	&(8.54$\pm$0.67)E12 &(7.22$\pm$0.39)E13\\ 
$-$16.1		&9.8	&(2.10$\pm$0.40)E12 &(1.21$\pm$0.15)E13\\ 
5.6		&10.3	&(2.65$\pm$0.38)E12 &(3.20$\pm$0.18)E13\\ 
34.2		&13.2	&(6.13$\pm$0.51)E12 &(5.36$\pm$0.26)E13\\ 
60.0		&6.8	&(3.13$\pm$0.44)E12 &(4.09$\pm$0.34)E13\\ 
77.0		&10.5	&(8.96$\pm$0.69)E12 &(9.18$\pm$0.54)E13\\ 
101.8		&3.5	&(1.26$\pm$0.29)E12 &(2.28$\pm$0.24)E13\\ 
116.9		&8.4	&(1.74$\pm$0.32)E12 &(5.06$\pm$0.29)E13\\
\hline 				       			 	 
\hline 				       			 	 
\end{tabular}			       			 	 
\end{center}			       			 	 
\end{table*}			       			 	 

In this system, we clearly detect the following species:  \aliii\ \l \l\ 1854, 1862 and \feii\ (\l \l\ 2260, 2344, 2374, 2382, 2586). However, \mnii\ (\l \l\ 2576, 2594, 2606); \tiii\ \l\ 1910; \crii\ \l\l\ 2056, 2062 and \znii\ \l \l\ 2026, 2062 are covered but not detected. Upper limits from non-detections implies log N(\mnii)$<$11.42, log N(\crii)$<$12.13 and log N(\znii)$<$11.76. The \mgi\ \l\ 2852 line fall in a spectral gap and the  \mgi\ \l\ 2026 line does not provide restrictive limits. The results of the profile fitting analysis are
summarised in Table~\ref{t:Q0132}, while Figure~\ref{f:Q0132} illustrates the fits.

\subsubsection{Q0138$-$0005; \lognhi=19.81; \zabs=0.7821; 5 components }  

\begin{figure*}
\begin{center}
\includegraphics[height=13cm, width=13cm, angle=0]{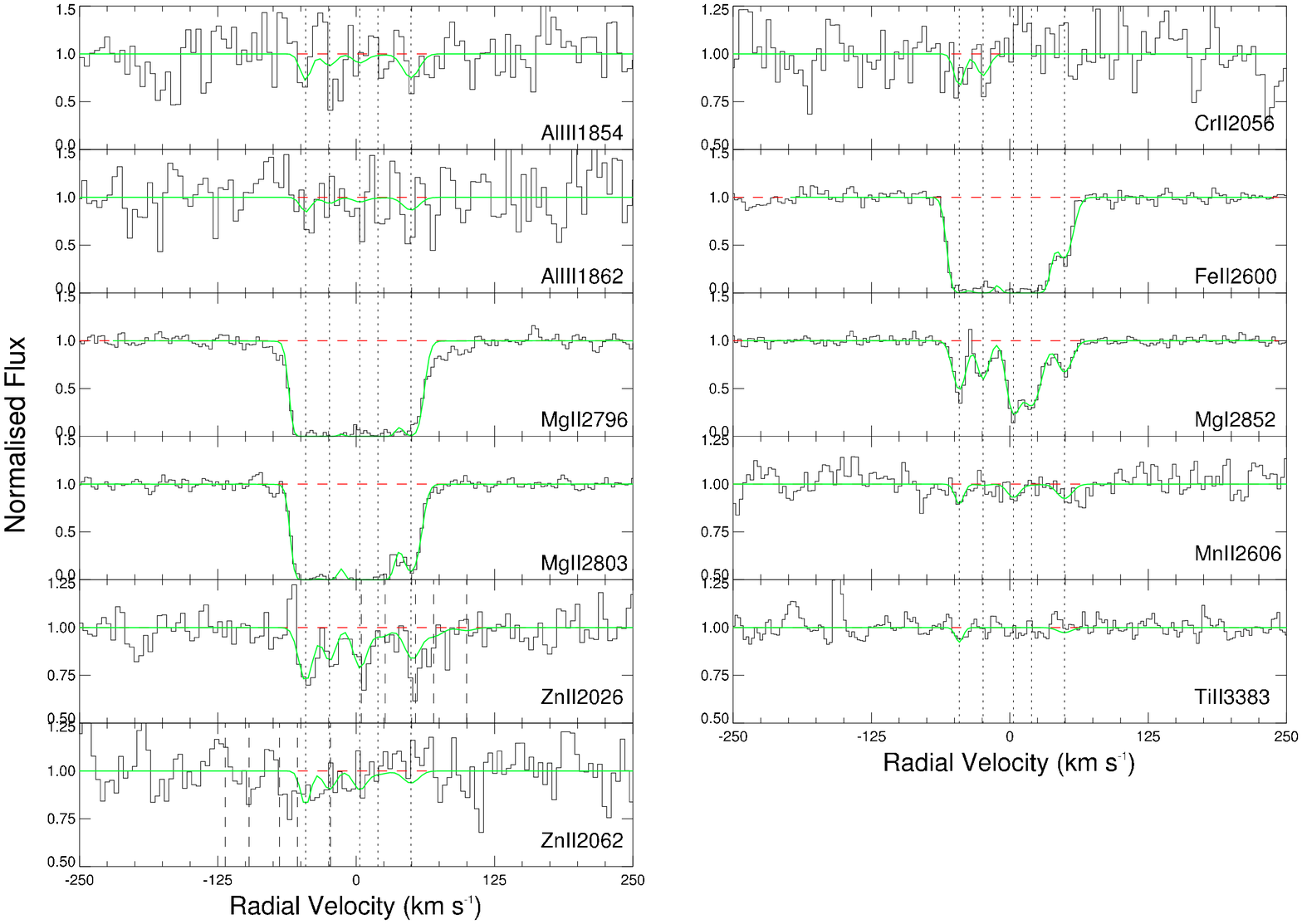}
\caption{Voigt profile fit of the \zabs=0.7821 sub-DLA towards Q0138$-$0005 (see Table~\ref{t:Q0138}).
\label{f:Q0138}}
\end{center}
\end{figure*}

\begin{table*}
\begin{center}
\caption{Parameters for the Voigt profile fit of the $z_{\rm abs} = 0.7821$ sub-DLA towards 
Q0138$-$0005 (see Figure~\ref{f:Q0138}). }
\label{t:Q0138}
\begin{tabular}{cccccccc}
\hline\hline
         Vel      & b    &N(\mgi) &N(\mgii) &N(\feii)   &N(\znii)   \\ 
 \hline 				       			 	 
$-$45.3	&7.0		&(7.29$\pm$0.50)E11	&	$>$2.20E14	&	$>$7.67E13	&	(1.89$\pm$0.22)E12	\\
$-$24.1	&6.1		&(4.68$\pm$0.40)E11	&	$>$8.59E13	&	$>$4.26E14	&	(1.00$\pm$0.19)E12	\\
3.3	&6.7		&(1.35$\pm$0.10)E12	&	$>$1.47E15	&	$>$6.95E14	&	(1.14$\pm$0.40)E12	\\
19.7	&9.9		&(1.54$\pm$0.09)E12	&	$>$1.32E14	&	$>$2.85E14	&	--			\\
49.4	&8.6		&(4.72$\pm$0.41)E11	&	$>$1.79E13	&	$>$1.03E13	&	(8.57$\pm$2.03)E11	\\
 \hline 				       			 	 
\hline 				       			 	 
\end{tabular}			       			 	 
\end{center}			       			 	 
\end{table*}

The following species are detected and fitted in this sub-DLA: \mgi\ \l\ 2852; \mgii\ \l \l\ 2796, 2803; and \znii\ \l \l\ 2026, 2062. Only the saturated \feii\ \l\ 2600 \AA\ line is detected (the other \feii\ lines fall in a spectral gap), the profile fittings of these lines provides lower limits to the column densities. The following lines: \aliii\ \l \l\ 1854, 1862; \siii\ \l\ 1808; \crii\ \l\ 2056; \mnii\ \l\ 2606 (the others lines for \mnii\ fall in a spectral gap); \tiii\ (\l \l\ 3073, 3230, 3242, 3384) and \nai\ (\l \l\ 3303, 3304) are covered but not detected. 
Upper limits from non-detections are log N(\aliii)$<$12.28, log N(\siii)$<$14.89, log N(\crii)$<$12.61, log N(\mnii)$<$11.84, log N(\tiii)$<$11.48 and log N(\nai)$<$ 13.05. The results of the profile fitting analysis are
summarised in Table~\ref{t:Q0138}, while Figure~\ref{f:Q0138} illustrates the fits.          
         
\subsubsection{Q0153$+$0009; \lognhi=19.70; \zabs=0.7714; 18 components }

\begin{figure*}
\begin{center}
\includegraphics[height=13cm, width=13cm, angle=0]{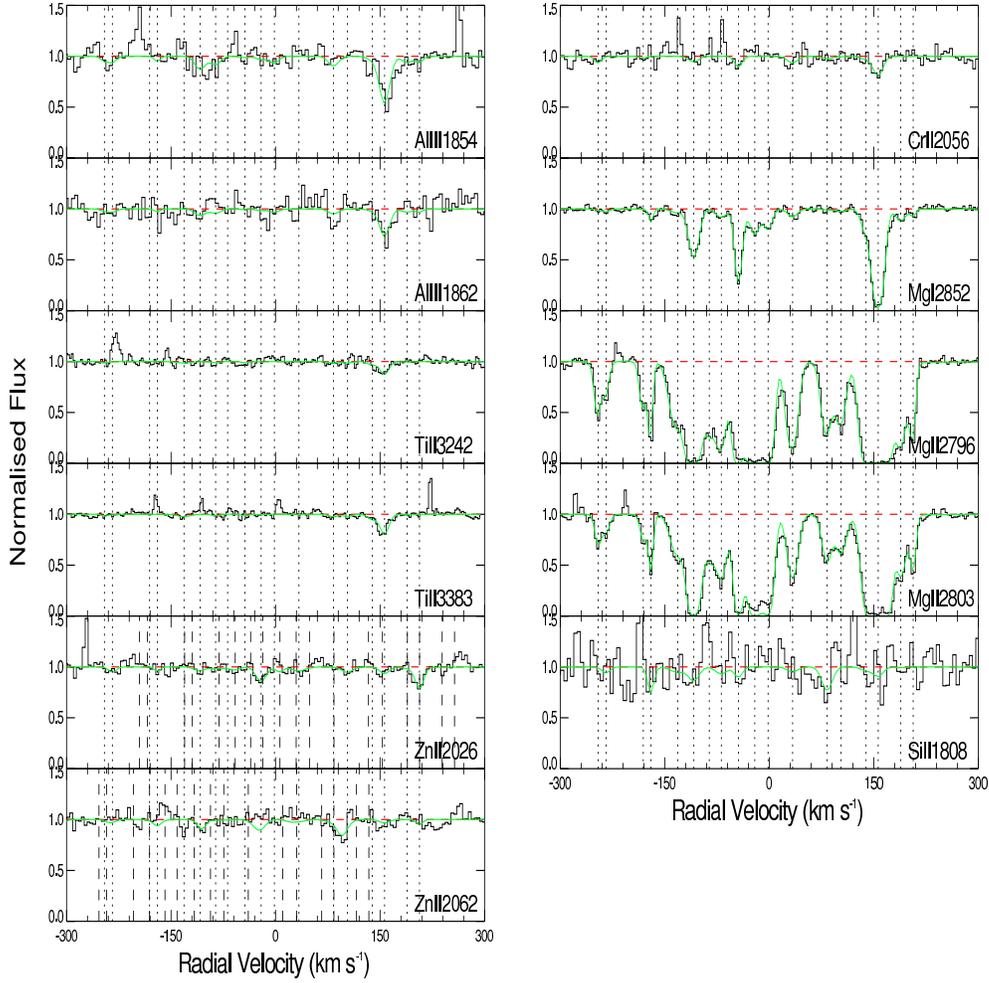}
\caption{Voigt profile fit of the \zabs=0.7714 sub-DLA towards Q0153$+$0009 (see Table~\ref{t:Q0153}).  Obvious apparent emission features are sky subtraction artifacts.
\label{f:Q0153}}
\end{center}
\end{figure*}

\begin{table*}
\begin{center}
\caption{Parameters for the Voigt profile fit of the $z_{\rm abs} = 0.7714$ sub-DLA towards Q0153$+$0009
(see Figure~\ref{f:Q0153}). Several lines of \feii\ are detected in the original SDSS spectra. The corresponding EW estimates are reported in Table~\ref{t:EW}.}
\label{t:Q0153}
\begin{tabular}{cccccccc}
\hline\hline
         Vel      & b    &N(\mgi) &N(\mgii) &N(\aliii)   &N(\crii)  &N(\tiii)   \\ 		      	
\hline
$-$245.7	&	5.2	&	--			&	$>$1.74E12	&	--			&	--			&	--	\\
$-$234.2	&	6.0	&	--			&	$>$1.11E12	&	--			&	--			&	--	\\
$-$180.8	&	4.9	&	--			&	$>$1.45E12	&	--			&	--			&	--	\\
$-$169.8	&	2.8	&	(7.69$\pm$1.28)E10	&	$>$3.43E12	&	--			&	--			&	--	\\
$-$131.4	&	12.8	&	(6.55$\pm$1.96)E10	&	$>$5.52E12	&	--			&	--			&	--	\\
$-$108.2	&	8.9	&	(8.04$\pm$0.27)E11	&	$>$4.01E13	&	-- 	&				&	-	\\
$-$86.1	&	8.7	&	-			&	$>$4.20E12	&	--			&		--		&	--	\\
$-$68.8	&	9.9	&	(1.17$\pm$0.18)E11	&	$>$8.19E12	&	--			&	--			&	--	\\
$-$44.1	&	5.9	&	(1.15$\pm$0.04)E12	&	$>$3.20E13	&	--			&	--			&	--	\\
$-$20.7	&	9.1	&	(3.60$\pm$0.20)E11	&	$>$5.19E13	&	--			&	--			&	--	\\
$-$1.7	&	7.2	&	(2.30$\pm$0.18)E11	&	$>$3.06E13	&	--			&	--			&	--	\\
33.5	&	9.3	&	(1.02$\pm$0.17)E11	&	$>$7.73E12	&	--			&	--			&	--	\\
83.1	&	8.6	&	-			&	$>$4.56E12	&	--			&	--			&	--	\\
103.1	&	9.3	&	(3.37$\pm$1.65)E10	&	$>$3.74E12	&	--			&	--			&	--	\\
138.9	&	8.6	&	(3.12$\pm$0.23)E11	&	$>$1.08E13	&	--			&	--			&	--	\\
156.4	&	8.3	&	(4.13$\pm$0.16)E12	&	$>$5.54E15	&	(4.00$\pm$0.50)E12	&	(1.04$\pm$0.13)E12	&	(6.50$\pm$1.52)E12	\\
188.8	&	9.4	&	(1.59$\pm$0.18)E11	&	$>$7.30E12	&	--			&	--			&	--	\\
206.6	&	4.3	&	(4.90$\pm$1.30)E10	&	$>$3.39E12	&	--			&	--			&	--	\\
 \hline 				       			 	 
\hline 				       			 	 
\end{tabular}			       			 	 
\end{center}			       			 	 
\end{table*}

The following species are clearly detected: \mgi\ \l\ 2852 and \mgii\ \l \l\ 2796, 2803. In addition, \aliii\ \l \l\ 1854, 1862; \crii\ \l\ 2056; and \tiii\ \l \l\ 3242, 3383 are all detected in the strongest component at v = 156.4 km s$^{-1}$. In this system, \mgi\ \l\ 2852 is also detected and is particularly saturated at $+$160 km s$^{-1}$. The \tiii\ lines are extremely strong, being at least 10 times stronger than in DLAs (Ledoux et al. 2002; Dessauges-Zavadsky et al. 2002), even though \nhi\  is 10 times weaker. \znii\ \l \l\ 2026, 2062; \siii\ \l\ 1808 and \nai\ (\l \l\ 3303, 3304) are covered but not detected. Upper limits from non-detections lead to log N(\znii)$<$11.96, log N(\siii)$<$14.49 and log N(\nai)$<$13.01. Unfortunately, all the \feii\ lines for that system fall in spectral gaps. However, several lines of \feii\ are detected in the original SDSS DR5 spectra. The results of the profile fitting analysis are summarised in Table~\ref{t:Q0153}, while Figure~\ref{f:Q0153} illustrates the fits.

\subsubsection{Q0449$-$1645; \lognhi=20.98; \zabs=1.0072; 13 components }  

\begin{figure*}
\begin{center}
\includegraphics[height=13cm, width=13cm, angle=0]{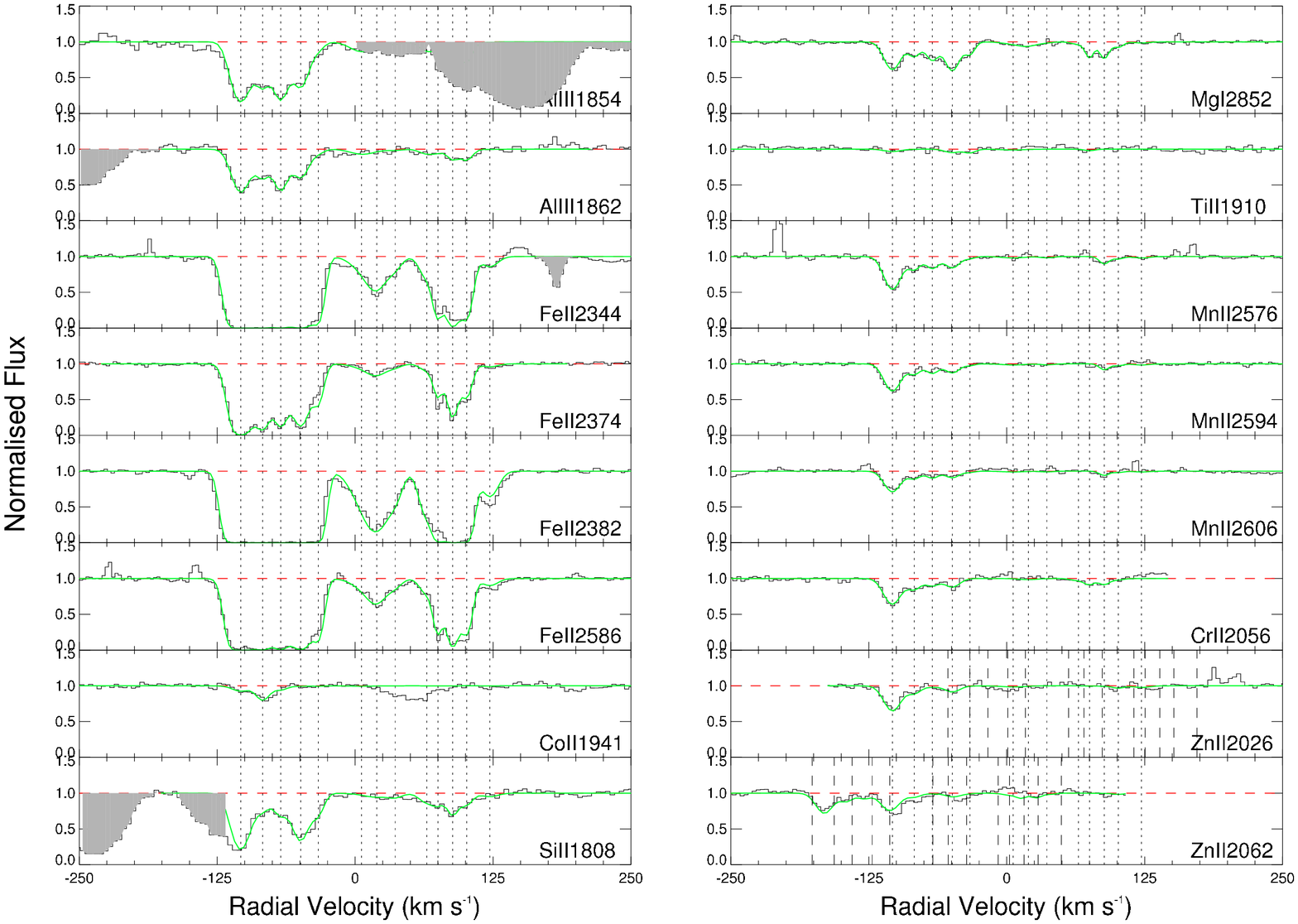}
\caption{Voigt profile fit of the \zabs=1.0072 DLA towards Q0449$-$1645 (see Table~\ref{t:Q0449}). The areas marked in grey are forest blends.
\label{f:Q0449}}
\end{center}
\end{figure*}

\begin{table*}
\begin{center}
\caption{Parameters for the Voigt profile fit of the $z_{\rm abs} = 1.0072$ DLA towards Q0449$-$1645 (see Figure~\ref{f:Q0449}).  \coii\ is detected in the first two components ($-$103.6 and $-$83.7 km s$^{-1}$) of the profile with (8.28$\pm$3.20)E12 and (1.90$\pm$0.33)E13 respectively. Confirmation with another \coii\ line is needed to show that this is not an interloper. }
\label{t:Q0449}
\begin{tabular}{ccccccccc}
\hline\hline
        Vel      & b    &N(\mgi) &N(\aliii)   &N(\siii)  &N(\crii) &N(\mnii) &N(\feii) &N(\znii)   \\ 		      	
\hline
$-$103.6	&	9.4	&	(6.26$\pm$0.20)E11	&	(1.28$\pm$0.07)E13	&	(3.00$\pm$0.14)E15	&	(1.48$\pm$0.08)E13	&	(4.45$\pm$0.36)E12	&	(4.86$\pm$0.26)E14	&		(2.99$\pm$0.08)E12	\\
$-$83.7	&	6.8	&	(2.16$\pm$0.14)E11	&	(5.44$\pm$0.50)E12	&	(4.12$\pm$1.85)E14	&	(4.22$\pm$0.58)E12	&	(1.10$\pm$0.23)E12	&	(1.78$\pm$0.15)E14	&		(6.04$\pm$0.53)E11	\\
$-$67.4	&	6.8	&	(2.85$\pm$0.15)E11	&	(9.43$\pm$0.68)E12	&	(5.52$\pm$2.01)E14	&	(2.80$\pm$0.55)E12	&	(1.00$\pm$0.23)E12	&	(1.49$\pm$0.10)E14	&		--	\\
$-$49.3	&	8.2	&	(5.82$\pm$0.20)E11	&	(6.37$\pm$0.51)E12	&	(1.92$\pm$0.39)E15	&	(3.80$\pm$0.59)E12	&	(1.11$\pm$0.24)E12	&	(1.69$\pm$0.08)E14	&		(5.55$\pm$0.55)E11	\\
$-$33.3	&	5.2	&	(1.47$\pm$0.13)E11	&	--			&	(4.60$\pm$1.91)E14	&	--			&	--			&	(4.80$\pm$0.25)E13	&		--	\\
5.8		&	13.6	&	--			&	(1.33$\pm$0.50)E12	&	--			&	--			&	--			&	(5.86$\pm$0.79)E12	&		--	\\
19.6		&	11.5	&	(8.21$\pm$1.93)E10	&	--			&	--			&	--			&	--			&	(1.79$\pm$0.10)E13	&		--	\\
36.3		&	7.7	&	--			&	--			&	--			&	--			&	--			&	(3.57$\pm$0.52)E12	&		--	\\
65.0		&	7.7	&	--			&	--			&	(2.34$\pm$1.75)E14			&	--			&	--			&	(7.67$\pm$0.62)E12	&  		--	\\
75.1		&	4.2	&	(1.68$\pm$0.13)E11	&	--			&	--			&	(1.67$\pm$0.51)E12	&	--			&	(3.22$\pm$0.19)E13	&		--	\\
88.4		&	4.9	&	(2.01$\pm$0.13)E11	&	(1.43$\pm$0.32)E12	&	(4.73$\pm$1.87)E14	&	(1.93$\pm$0.49)E12	&	(4.84$\pm$1.88)E11	&	(8.20$\pm$0.45)E13	&		--	\\
101.1	&	5.3	&	(6.31$\pm$1.14)E10	&	(1.57$\pm$0.32)E12	&	(2.38$\pm$1.53)E14			&	--			&	--			&	(3.82$\pm$0.18)E13	&		--	\\
122.1	&	9.3	&	--			&	--			&	--			&	--			&	--			&	(3.89$\pm$0.48)E12	&		--	\\
\hline
\hline 				       			 	 
\end{tabular}			       			 	 
\end{center}			       			 	 
\end{table*}			       			 	 
         
The following features are detected and fitted: \mgi\ \l\ 2852; \aliii\ \l \l\ 1854 (partly blended with Lyman-$\alpha$ forest features) and 1862 (clean); \siii\ \l\ 1808; \crii\ \l\ 2056; \mnii\ (\l \l\ 2576, 2594, 2606); \feii\ (\l \l\ 2344, 2374, 2382, 2586); \znii\ \l \l\ 2026, 2062; and \coii\ \l\ 1941. Lines of \tiii\ (\l \l\  1910, 3073, 3230, 3242, 3384) and \nai\ (\l \l\ 3303, 3304) are covered but not detected. 
Upper limits from non-detections lead to log N(\tiii)$<$11.47 and log N(\nai)$<$12.66. The \mgii\ doublet for that systems falls in a spectral gap.  Interestingly, the \coii\ is mutually exclusive of the \siii. The \siii\ mimics other elements, but for instance, the strength ratio [\siii/\mnii] is quite different between $-$104 km s$^{-1}$and  $-$49 km s$^{-1}$. Confirmation with another \coii\ line is needed to show that this is not an interloper. The results of the profile fitting analysis are summarised in Table~\ref{t:Q0449}, while Figure~\ref{f:Q0449} illustrates the fits. The areas marked in grey are blended with Lyman-$\alpha$ forest features.

\subsubsection{Q2335$+$1501; \lognhi=19.70; \zabs=0.6798; 6 components} 

\begin{figure*}
\begin{center}
\includegraphics[height=13cm, width=13cm, angle=0]{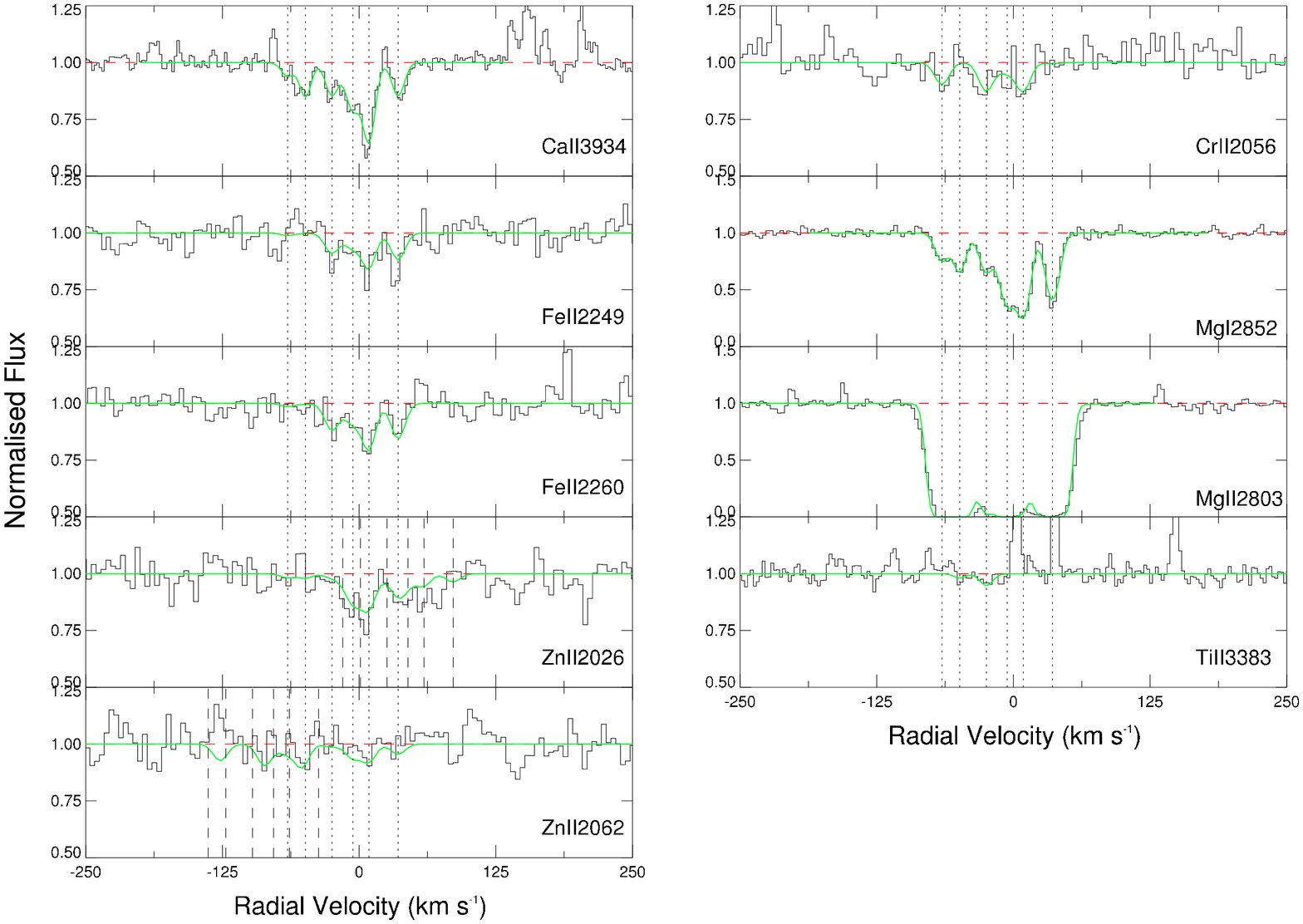}
\caption{Voigt profile fit of the \zabs=0.6798 sub-DLA towards Q2335$+$1501 (see Table~\ref{t:Q2335}).
\label{f:Q2335}}
\end{center}
\end{figure*}

\begin{table*}
\begin{center}
\caption{Parameters for the Voigt profile fit of the $z_{\rm abs} = 0.6798$ sub-DLA towards Q2335$+$1501 (see Figure~\ref{f:Q2335}).}
\label{t:Q2335}
\begin{tabular}{cccccccccc}
\hline\hline
          Vel      & b    &N(\mgi) &N(\mgii)   &N(\caii)  &N(\crii) &N(\feii &N(\znii)   \\ 		      	
\hline
$-$65.3	&	7.5	&	(3.07$\pm$0.21)E11	&	$>$2.32E14	&	(1.36$\pm$0.63)E11	&	--			&	--			&	--	\\
$-$49.1	&	6.8	&	(4.29$\pm$0.23)E11	&	$>$3.77E14	&	(3.26$\pm$0.65)E11	&	--			&	--			&	--	\\
$-$24.8	&	7.8	&	(4.75$\pm$0.24)E11	&	$>$2.42E13	&	(3.49$\pm$0.69)E11	&	(4.02$\pm$1.33)E12	&	(1.40$\pm$0.27)E14	&	--	\\
$-$5.6	&	8.2	&	(1.19$\pm$0.05)E12	&	$>$1.07E14	&	(5.53$\pm$0.79)E11	&	--			&	(1.08$\pm$0.28)E14	&	(8.40$\pm$1.44)E11	\\
9.0	&	7.2	&	(1.41$\pm$0.06)E12	&	$>$1.68E13	&	(8.84$\pm$0.86)E11	&	(3.81$\pm$1.41)E12	&	(2.39$\pm$0.29)E14	&	(9.43$\pm$1.39)E11	\\
35.7	&	7.5	&	(9.80$\pm$0.04)E11	&	$>$2.04E15	&	(3.43$\pm$0.46)E11	&	--			&	(1.85$\pm$0.27)E14	&	(5.57$\pm$1.27)E11	\\
 \hline 				       			 	 
\hline 				       			 	 
\end{tabular}			       			 	 
\end{center}			       			 	 
\end{table*}			       			 	 
         
We report the detection of the following features: \mgi\ \l\ 2852; \mgii\ \l\ 2803; \crii\ \l\ 2056; \feii\ \l \l\ 2249, 2260; \znii\ \l \l\ 2026, 2062 and \caii\ \l\ 3934. The \mgii\ \l\ 2796 is falling right at the peak of an emission line thus rendering continuum fitting nontrivial. \tiii\ (\l \l\ 3073, 3230, 3242, 3384) and \nai\ (\l \l\ 3303, 3304) are covered but not detected. Upper limits from non-detections lead to log N(\tiii)$<$11.36 and log N(\nai)$<$12.75. The \mnii\ \l\l\l\ 2576, 2594, 2606 lines for that systems fall in a spectral gap. The results of the profile fitting analysis are summarised in Table~\ref{t:Q2335}, while Figure~\ref{f:Q2335} illustrates the fits.

\subsection{Results: Equivalent Widths, Total Column Densities and Abundances}

\begin{table*}
\footnotesize
\caption{Rest frame equivalent widths and associated errors for the systems in this sample. Values and 3-$\sigma$ errors are in units of m\mbox{\AA}. The EW of \tiii\ \l\ 3242 in the sub-DLAs towards Q0153$+$0009 is 27$\pm$7 m\AA. Numbers in brackets are from the original SDSS spectra with 1-$\sigma$ errors (York et al. 2005; York et al. 2006).}
\label{t:EW}
\begin{tabular}{ccccccccccccc}
\hline\hline
		&log 	&	\mgi\		 	&		\mgii\		 	&	\mgii\ 		 	&	\aliii\ 		 	&	\aliii\	 	&	\siii\ 		 	&	\caii\ 		 		&	\tiii\			&	\znii$^a$   &\znii$^b$			\\
		&\hi\ 		&	2852			&	2796			&	2803			&	1854			&	1862			&	1808			&	3933			&	3383			&	2026			&	2062			\\
\hline                                                                                                                                                                                                                                                                                                                                                          
Q0123$-$0058	&20.08 &(799$\pm$95)			&	(1911$\pm$75)			&	(1895$\pm$78)			&	616$\pm$34		&	297$\pm$31		&	154$\pm$27		&	-			&	-			&	$<$7			&	$<$10			\\
Q0132$+$0116 &19.70	&-			&	-			&	-			&	469$\pm$35		&	318$\pm$41		&	-			&	-			&	-			&	$<$7			&	$<$7			\\
Q0138$-$0005	&19.81 &408$\pm$13		&	1186$\pm$15		&	1065$\pm$10		&	$<$40			&	$<$44			&	$<$75			&	(325$\pm$59)			&	$<$7			&	131$\pm$31		&	$<$19			\\
Q0153$+$0009 &19.70	&559$\pm$34		&	2563$\pm$22		&	2058$\pm$24		&	84$\pm$25		&	33$\pm$13		&	$<$27			&	(319$\pm$60)			&	46$\pm$11		&	$<$7			&	$<$11			\\
Q0449$-$1645&20.98 	&287$\pm$16		&	-			&	-			&	346$\pm$9		&	270$\pm$15		&	257$\pm$20		&	-			&	-			&	33$\pm$6		&	160$\pm$15		\\
Q2335$+$1501 &19.70	&432$\pm$10		&	974$\pm$11		&	1268$\pm$8		&	-			&	-			&	-			&	193$\pm$25		&	$<$7			&	81$\pm$25		&	$<$12			\\
\hline\hline																																															
	
\hline\hline																																														
	&z$_{\rm abs}$	&	\mnii\ 		 	&	\mnii\ 		 	&	\mnii\		 	&	\feii\ 		 	&	\feii\ 		 	&	\feii\		 	&	\feii\	 	&	\feii\		 	&	\feii\ 		 	&	\crii\ 						\\
		 &		&	2576			&	2594			&	2606			&	2260			&	2344			&	2374			&	2382			&	2586			&	2600			&	2056		 	\\
\hline                                                                                                                                                                                                                                                                                                                                                         
Q0123$-$0058 &	1.4094	&	74$\pm$13		&	-			&	-			&	118$\pm$34		&	1086$\pm$31		&	740$\pm$21		&	1247$\pm$27		&	-			&	-			&	$<$10			\\
Q0132$+$0116 &	1.2712	&	$<$8			&	$<$6			&	$<$7			&	$<$7			&	1184$\pm$23		&	518$\pm$31		&	1744$\pm$20		&	1023$\pm$26		&	1782$\pm$20		&	$<$6			\\
Q0138$-$0005 &	0.7821	&	-			&	-			&	$<$11			&	-			&	-			&	-			&	-			&	-			&	898$\pm$12		&	$<$20			\\
Q0153$+$0009 &	0.7714	&	-			&	-			&	-			&	-			&	-			&	(580$\pm$87)			&	(1258$\pm$77)		&	(944$\pm$76)			&	(1326$\pm$73)			&	21$\pm$5		\\
Q0449$-$1645 &	1.0072	&	154$\pm$11		&	125$\pm$21		&	64$\pm$16		&	-			&	1156$\pm$10		&	803$\pm$11		&	1465$\pm$9		&	1174$\pm$12		&	1559$\pm$12		&	93$\pm$18		\\
Q2335$+$1501 &	0.6798	&	(295$\pm$84)			&	-			&	-			&	56$\pm$11		&	-			&	-			&	-			&	-			&	-			&	$<$11			\\
\hline\hline
	\end{tabular}
\vspace{0.2cm}
\begin{minipage}{140mm}
{\bf $^a$:} the contribution of \mgi\ \l\ 2026 is not taken into account in the estimate of the EW of \znii\ \l\ 2026. \\
{\bf $^b$:} the contribution of \crii\ \l\ 2062 is not taken into account in the estimate of the EW of \znii\ \l\ 2062. \\
\end{minipage}
\end{table*}

For each transition, we determine the rest-frame equivalent width (EW) value or limit at the expected position of the line. The 3-$\sigma$ errors for the equivalent widths reflect both uncertainties in the continuum level and in the photon noise. If a certain line was not detected, the limiting equivalent widths are 3-$\sigma$ limits, based on a 3-pixel wide line (after rebinning to 0.05\AA/pix). Results from this study are shown in Table~\ref{t:EW}, numbers in parentheses are from the original SDSS spectra with 1-$\sigma$ errors (York et al. 2005; York et al. 2006). When overlapping, the two are broadly in agreement.

\begin{table*}
\begin{center}
\caption{Logarithmic values of the total column densities for each absorber. For each entry, the first line refers to Voigt profile fitting and the second to the Apparent Optical Depth (AOD) technique. Co is detected towards Q0449$-$1645 and Ca towards Q2335$+$1501. Transitions marked "--" are not covered by our spectra.}
\label{t:total_N}
\begin{tabular}{cccccccc}
\hline\hline
                &\lognhi             & \zabs &\mgi           &\mgii              &\aliii          &\siii          &\crii  \\
\hline
Q0123$-$0058  	&	20.08$^{+0.10}_{-0.08}$   &	1.4094	&	--             	&	--       	&	13.76$\pm$0.02 	&	15.47$\pm$0.05 	&	12.91$\pm$0.10  \\
AOD		&				&		&	--		&	--		&	13.67$\pm$0.05	&	15.50$\pm$0.07	&	12.88$\pm$0.15 \\
Q0132$+$0116  	&	19.70$^{+0.08}_{-0.10}$   &	1.2712	&	--             	&	--       	&	13.58$\pm$0.02 	&	--             	&	$<$12.13$^a$    \\  
AOD		&				&		&	--		&	--		&	13.55$\pm$0.03	&	--		&	--		\\
Q0138$-$0005  	&	19.81$^{+0.06}_{-0.11}$   &	0.7821	&	12.66$\pm$0.01 	&	$>$15.57 	&	$<$12.28$^a$   	&	$<$14.89$^a$   	&	$<$12.61$^a$    \\
AOD		&				&		&	12.66$\pm$0.01 	&	$>$14.29	&	--		&	--		&	--		\\
Q0153$+$0009  	&	19.70$^{+0.08}_{-0.10}$   &	0.7714	&	12.88$\pm$0.01 	&	$>$15.76 	     &	12.60$\pm$0.XX 	&	$<$14.49$^a$   	&	12.81$\pm$0.10    \\ 
AOD		&				&		&	12.84$\pm$0.02	&	$>$14.36	&	12.77$\pm$0.11	&	--		&	12.78$\pm$0.08 \\
Q0449$-$1645  	&	20.98$^{+0.06}_{-0.07}$   &	1.0072	&	12.37$\pm$0.01 	&	--       	&	13.58$\pm$0.02	&	15.86$\pm$0.04 	&	13.47$\pm$0.02  \\
AOD		&				&		&	12.40$\pm$0.02	&	--		&	13.61$\pm$0.02	&	15.73$\pm$0.03	&	13.44$\pm$0.08 \\
Q2335$+$1501  	&	19.70$^{+0.30}_{-0.30}$ 	&	0.6798	&	12.68$\pm$0.01 	&	$>$15.45 &	--             	&	--             	&	12.89$\pm$0.10  \\
AOD		&				&		&	12.67$\pm$0.01	&	$>$14.39	&	--	&	--		&	13.04$\pm$0.22 \\
\hline
\hline
\end{tabular}			       			 	 
\begin{tabular}{ccccccccc}
\hline\hline
                &\lognhi             & \zabs &\mnii           &\feii          &\znii                 &\tiii          &\coii\ or \caii         &\nai             \\
\hline
Q0123$-$0058  	&	20.08$^{+0.10}_{-0.08}$   &	1.4094	&	12.61$\pm$0.12 	&	14.98$\pm$0.02 	&	12.23$\pm$0.10	&	--             	&	--             	&	--                \\
AOD		&				&		&	12.62$\pm$0.06	&	15.08$\pm$0.12	&	12.41$\pm$0.19	&	--		&	--		&	--		\\
Q0132$+$0116  	&	19.70$^{+0.08}_{-0.10}$   &	1.2712	&	$<$11.42$^a$   	&	14.62$\pm$0.01 	&	$<$11.76$^a$   	&	--             	&	--             	&	--                \\
AOD		&				&		&	--		&	14.62$\pm$0.03	&	--		&	--		&	--		&	--		\\
Q0138$-$0005  	&	19.81$^{+0.06}_{-0.11}$   &	0.7821	&	$<$11.84$^a$   	&	$>$15.17       	&	12.69$\pm$0.05 	&	$<$11.48$^a$   	&	--             	&	$<$13.05$^a$        \\
AOD		&				&		&	--		&	$>$14.32	&	12.91$\pm$0.10	&	--		&	--		&	--                  \\
Q0153$+$0009  	&	19.70$^{+0.08}_{-0.10}$   &	0.7714	&	--             	&	--             	&	$<$11.96$^a$   	&	12.02$\pm$0.06 	&	--             	&	$<$13.01$^a$       \\
AOD		&				&		&	--		&	--		&	--		&	12.13$\pm$0.09	&	--		&	--		\\
Q0449$-$1645  	&	20.98$^{+0.06}_{-0.07}$   &	1.0072	&	12.91$\pm$0.03 	&	15.09$\pm$0.01 	&	12.62$\pm$0.07 	&	$<$11.47$^a$   	&	13.51$\pm$0.07 	&	$<$12.66$^a$      \\
AOD		&				&		&	12.92$\pm$0.03	&	15.07$\pm$0.03	&	12.32$\pm$0.08	&	--		&	13.49$\pm$0.04	&	--		\\
Q2335$+$1501  	&	19.70$^{+0.30}_{-0.30}$	&	0.6798	&	--             	&	14.83$\pm$0.03 	&	12.37$\pm$0.04 	&	$<$11.36$^a$   	&	12.41$\pm$0.03 	&	$<$12.75$^a$       \\
		&				&		&	--		&	14.74$\pm$0.08	&	12.69$\pm$0.13	&	--		&	12.40$\pm$0.05	&	--		\\
\hline
\hline 				       			 	 
\end{tabular}			       			 	 
\end{center}			       			 	 
\vspace{0.2cm}
\begin{minipage}{140mm}
{\bf $^a$:} 3-$\sigma$ limit calculated from the non-detection of the line.\\
\end{minipage}
\end{table*}			       			 	 

Total column densities for all observed species are presented in Table~\ref{t:total_N} for each of the six absorption systems studied here. For each system, the top line presents the sum of the column densities obtained from the Voigt profile fits while the column densities derived from Apparent Optical Depth (AOD) method (Savage \& Sembach 1991) are shown on the line below. In all cases, both methods give consistent results within the error estimates. We then estimate the abundances of elements with respect to solar for each system following the standard definition:

\begin{equation}
[X/Y] =\log [N(X)/N(Y)]_{absorber}- \log [N(X)/N(Y)]_{\odot}
\label{eq}
\end{equation}

where $\log [N(X)/N(Y)]_{\odot}$ is the logarithmic solar abundance and is taken
from Asplund et al. (2005) adopting the mean of photospheric and
meteoritic values for Mg, Si, Mn, Fe, Zn, Ti, Co and Ca. These values are
recalled on the top line of Table~\ref{t:ab}. The last column of the table summarises the $\Delta (g-i)$ for 
each absorber, where $\Delta (g-i)$ is defined as:

\begin{equation}
\Delta (g-i)=(g-i)-(g-i)_{\rm med}
\end{equation}

where $(g-i)_{\rm med}$ is the median (g-i) colour of a control sample
of 100 quasars with the closest magnitude to the object within a redshift
interval of $\pm$ 0.05. The resulting error estimate is the colour dispersion of this control sample combined with the error in the colour measure as done in Vladilo et al. (2006). $\Delta (g-i)$ colours are calculated from SDSS DR5 release photometric values. The object Q0449$-$1645 is not a SDSS quasar and therefore no $\Delta (g-i)$ is available for this system. For Q0132$+$0116, SDSS DR3 release photometric values were used, since the DR5 release values are not available. The uncertainty in $\Delta$(g-i) is large (see Table\ref{t:ab}), so that 0 reddening systems could have values of $\Delta$(g-i) between $-$0.2 and $+$0.2 (York et al. 2006).

\section{Discussion}

\subsection{Metal-Rich Systems}

In the sample presented here, we find two new {\it solar or above solar} sub-DLAs towards Q0138$-$0005 with [Zn/H]=$+$0.28$\pm$0.16 with rather high metallicity in iron too: [Fe/H]$>-$0.09 (i.e. Fe could be depleted compared to Zn by a factor of 2 or more) and towards Q2335$+$1501 with [Zn/H]=$+$0.07$\pm$0.34. One of these high Zn abundance systems (Q0138$-$0005) has large $\Delta$(g-i)=$+$0.45$\pm$0.13, and thus seems to be dusty, but not to have Fe significantly depleted. The elements Si, Cr, Mn and Ti are depleted compared to Zn. There could always be extinction arising in the circum quasar region which is hot and therefore does not show absorption lines. Furthermore, there could be another absorption system at lower redshift that is causing the  extinction. However, in the case of Q0138$-$0005, we have checked that the SDSS spectrum for this object shows no Broad Absorption Line (BAL) feature and no identified narrow line systems other than the target sub-DLA. We have checked that there are no \caii\ at z$<$0.7 and no \nai\  at z$<$0.4. The 1-$\sigma$ EW limit of detection for \nai\ is 0.27 \AA\ and for \caii, it is 0.15 to 0.2 \AA.

It is interesting to note that the situation with Fe and Si is reversed from the case of the SMC, where some sightlines have Si undepleted but Fe depleted (e.g. Welty 2003). Also, some of the dust might not be made of Fe, thus explaining the $\Delta$(g-i) reported here (but see also Vladilo 2002; Sofia et al. 2006; Li, Miseelt \& Wang 2006). 

The sub-DLAs in this study may need some ionisation corrections: as shown by Meiring et al. (2008), the corrections for stronger ionising radiation fields may lead to Fe depletion compared to Zn and Cr, but should not affect Ti, which has an ionisation potential near that of \hi. Indeed, simulation with the CLOUDY photo-ionisation model shows that [Zn/Cr] is affected by ionisation. Much of the correction needed is Zn though, which is the most affected element of the two. But this could be partly due to uncertain atomic data for Zn, and in any case ionisation corrections for Zn lead to higher Zn abundances and higher [Zn/Cr] (e.g. Vladilo et al. 2001; Meiring et al. 2007).

One more absorber may be dust free and have relatively high [Zn/H], Q0123$-$0058 with [Zn/H]=$-$0.45$\pm$0.20, while two systems with upper limits on Zn could have higher [Zn/H] than most DLAs, but new data are needed to confirm this: Q0132$+$0116 with [Zn/H]$<-$0.54 and Q0153$+$0009 with [Zn/H]$<-$0.34 compared to the DLA (Q0449$-$1645 with [Zn/H]=$-$0.96$\pm$0.08) and to DLAs in general (Kulkarni et al. 2007). These results are therefore in line with our recent findings (P\'eroux et al. 2006a; Kulkarni et al. 2007; Khare et al. 2007) that sub-DLAs show higher metallicities at low-redshift than classical DLAs. These systems may represent the
tip of the iceberg of a metal-rich population of higher \hi\ column
density systems currently hidden by dust extinction effects. If there really are very few DLAs with solar metallicity, then DLAs may mainly come from dwarf galaxies (Khare et al. 2007). In any case, the high abundance sub-DLAs are comparable with the ones measured in actively star-forming galaxies such as the
Lyman-Break Galaxies at higher redshifts (Pettini et al. 2002) and therefore offer a crucial
bridge between galaxies studied in emission and galaxies studied in absorption.

\subsection{Dust Content}

\begin{figure}
\begin{center}
\includegraphics[height=8cm, width=8cm, angle=0]{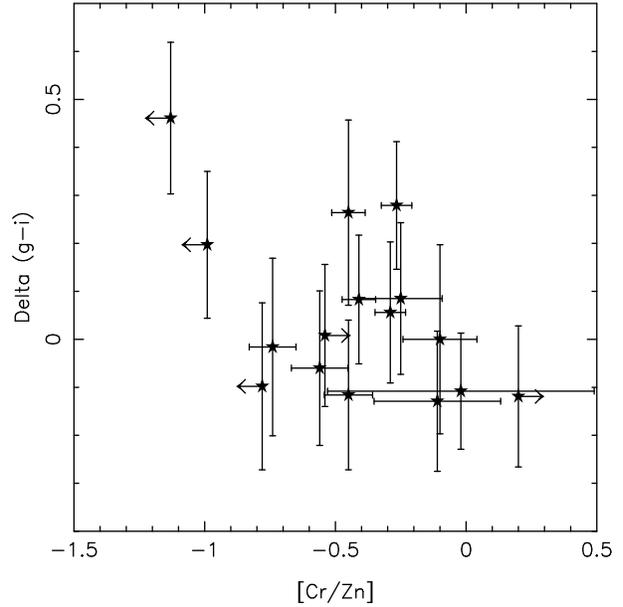}
\caption{$\Delta$(g-i) versus [Cr/Zn] for sub-DLAs and DLAs from this study and from the literature (Meiring et al. 2007; 2008; Prochaska et al. 2006; P\'eroux et al. 2006a; 2006b; Vladilo et al. 2006). The stars are measurements or upper/lower limits (as indicated by the arrows). There is still too few measurements to show a strong relation between these ratios and the amount of reddening measured by $\Delta$(g-i).
\label{f:ZnCr_Dgi}}
\end{center}
\end{figure}

As opposed to [Zn/H], most of the [Fe/H] measurements presented here are sub-solar (only Q0138$-$0005 has [Fe/H]$>-$0.09, i.e. not definitely sub-solar). The ratio of depleted (e.g., Cr, Fe) to undepleted
(Zn) elements provides a measure of the dust content of quasar absorbers (e.g. Pettini et al. 1997). The larger
the deficiency relative to the solar ratio [Cr/Zn]=0, the larger is the
fraction of chromium depleted from the gas to the dust phase. The refractory elements such as Cr and Fe might be expected to show a correlation between
their abundances relative to the mildly depleted elements such as Zn and the inferred reddening. We find [Cr/Zn]=$-$0.35 in Q0123$-$0058, [Cr/Zn]$<-$0.42 in Q0138$-$0005, [Cr/Zn]$>-$0.18 in Q0153$+$0009 and [Cr/Zn]=$-$0.51 in Q2335$+$1501. Finally [Cr/Zn]=$-$0.18 in the DLA towards Q0449$-$1645.  There are still too few measurements to show a strong relation between these ratios and the amount of reddening measured by $\Delta$(g-i). Indeed, as shown in Fig~\ref{f:ZnCr_Dgi}, we observe large depletions as measured by [Cr/Zn] being
associated with both positive (Q0123$-$0058) and negative (Q2335$+$1501) values of
$\Delta$(g-i).  In this redshift regime, E(B-V) $\sim$ 0.25 $\times$ $\Delta$(g-i): the extinction in the three systems with the lowest $\Delta$(g-i) may be $<$ 0.05 in E(B-V) but the depletions can still be significant (e.g. York and Kinahan 1979). The only DLA in our sample towards Q0449$-$1645 seems to have low extinction, consistent with observed trends for zinc in DLAs. This is consistent with findings that DLAs show little depletion while sub-DLAs show considerable depletion at times (e.g. Table 12 of Meiring et al. 2007; Table 14 of Meiring et al. 2008). In fact, if dust extinction in quasar absorbers is a strong function of N($\rm Zn~{\sc ii}$), the obscuration will start acting at a lower [Zn/H] ratio for DLAs than for sub-DLAs (Vladilo \& P\'eroux 2005). In other words, we might be missing relatively more systems in the DLA range than in the sub-DLA range. Then, the fact that the average metallicity in observed DLAs is lower than in observed sub-DLAs would be partly explained.

\subsection{Abundances}

{\bf Calcium II:} \caii\ H and K lines might constitute a secure tool with which to detect DLA at
low-redshift even from medium-resolution optical spectra. Wild, Hewett \&
Pettini (2006) show that the 37 \caii\ absorbers found in SDSS, are remarkably dusty.
They also report a clear correlation between the strength of the \caii\ equivalent width and the amount of reddening in the background quasar spectra. These findings are surprising given that \caii\ is known to be highly
depleted onto dust grains and that, therefore, one would expect dusty
absorbers to have small {\it measured} gas phase \caii\ abundances. To better
understand the relation between observed \caii\ and dust,
higher-resolution data of quasar absorbers with \nhi\ measured where
the dust content can be estimated with other means (i.e. \crii/\znii)
are required. Here, we report a new high-resolution measure of CaII column density in an absorber at \zabs=0.6798 towards Q2335$+$1501. We derive log N(\caii)=12.41$\pm$0.03. Given that some \caiii\ might also be present in \hi\ region, this leads to an upper limit: [Zn/Ca]$\la$1.67 (while [Zn/Cr]=$+$0.54 and [Zn/Fe]=$+$0.39). However, Ca is not be a reliable indicator of the presence or absence of dust in the disk. Indeed, if dust is present anywhere in the disk, the Routly-Spitzer effect (Routly \& Spitzer 1952; see below) could still produce high [Ca/H] in some of the gas.

{\bf Cobalt II:} Almost no Co is found in the ISM, so the
reference standard is tough to set. Based on the results of Mullman et al. (1998), Co may be somewhat less depleted in the Galaxy than Fe (perhaps similar to Cr in cold clouds,
slightly less depleted in warm clouds). Ellison et al. (2001b) reported the first detection of cobalt in a quasar absorber with [Co/H]=$-$0.50 and [Co/Fe]=$+$0.31$\pm$0.05. Since then, other detections have been reported by Rao et al. (2005) with [Co/Fe]=$+$0.05$\pm$0.12 and Dessauges-Zavadsky et al. (2006) with [Co/Fe]=$+$0.24$\pm$0.13. Here, we present a new high-resolution detection of cobalt in the DLA towards Q0449$-$1645 with log N(\coii)=13.51$\pm$0.07 leading to [Co/H]=$-$0.36$\pm$0.14 and [Co/Fe]=$+$0.98. This value is quite high, but it is not excluded that an interloper is responsible for the absorption instead of \coii.

{\bf Titanium II:} The few abundance measures of Ti to date in DLAs suggest that the relative abundance
of Ti with respect to Fe is perhaps slightly super-solar. Since stellar Ti abundances behave as an alpha-capture element (such as Mg, Si, S) in the Galaxy, this may indicate both low dust depletion and/or be
evidence for an alpha-product enhancement for Type-II SNe. Ti is normally a highly depleted element, so high [Ti/Zn] means low dust. Furthermore, the interpretation is mitigated by the Routly-Spitzer effect where Fe in dust grains might be released in the high velocity clouds and strengthen the wings of \feii\ relative to \mgii\ (York et al. 2006). Thus far, only a few Ti measurements exist in DLAs at $z_{\rm abs} < 1.5$ (Ledoux et al. 2002; Dessauges-Zavadsky et al. 2002), while efforts to measure Ti in DLAs at $z_{\rm abs} > 1.5$ have focused on the pair of weak lines at $\lambda_{\rm rest} \approx 1910$ \AA\ and have yielded only a few tenuous detections (e.g. Prochaska \& Wolfe 1997, 1999; Prochaska et al. 2001). Our observations of the stronger Ti II $\lambda$3073, 3242, 3384 lines have resulted in one robust detection and two significant upper limits. We derive [Ti/H]=$-$0.57$\pm$0.16 in Q0153$+$0009;  [Ti/H]$<-$1.22 in Q0138$-$0005; [Ti/H]$<-$2.11 in Q0449$-$1645 and [Ti/H]$<-$1.23 in Q2335$+$1501. The range ratios N(\tiii)/N(\hi) is consistent with the range in diffuse interstellar clouds: the depletion is as much as  a factor of 100, in our case.

{\bf Sodium I:} Finally,  we note that while lines of \nai\ were covered in several lines of sight, we
did not detect the \l\ 3302 doublet absorption in any system. This is not surprising, given the
general weakness of these features in low \hi\ column lines-of-sight in the Galaxy.

\begin{table*}
\begin{center}
\caption{Abundances with respect to solar, [X/H]. The error bars on [X/H] include both the errors in
log $N(X)$ and \lognhi. Co is detected towards Q0449$-$1645 and Ca towards Q2335$+$1501 (the abundance is derived from \caii\ only). Transitions marked "--" are not covered by the current spectra.}
\label{t:ab}
\begin{tabular}{cccccccccc}
\hline\hline
             &Mg 	&Si 		  &Cr 		    &Mn     	      &Fe 		&Zn               &Ti              &Co/Ca 	    &$\Delta$(g-i)\\
\hline
A(X/H)$_{\sun}$ &$-$4.47 &$-$4.46          &$-$6.37          &$-$6.53          &$-$4.55          &$-$7.40          &$-$7.11         &$-$7.11/$-$5.69 &--\\
\hline
Q0123$-$0058 &--	&$-$0.15$\pm$0.15 &$-$0.80$\pm$0.20 &$-$0.94$\pm$0.21 &$-$0.55$\pm$0.12 &$-$0.45$\pm$0.20&--               &--             &$+$0.16$\pm$0.16\\
Q0132$+$0116 &--	&--		  &$<-$1.20$^a$	    &$<-$1.75$^a$     &$-$0.53$\pm$0.12	&$<-$0.54$^a$	  &--	  	    &--	  	    &$-$0.07$\pm$0.12\\
Q0138$-$0005 &$>+$0.23  &$<-$0.46$^a$	  &$<-$0.83$^a$	    &$<-$1.44$^a$     &$>-$0.09		&$+$0.28$\pm$0.16 &$<-$1.22$^a$     &--	            &$+$0.45$\pm$0.13\\
Q0153$+$0009 &$>+$0.53  &$<-$0.75$^a$	  &$-$0.52$\pm$0.20 &--		      &--		&$<-$0.34$^a$	  &$-$0.57$\pm$0.16 &--		    &$-$0.13$\pm$0.15\\
Q0449$-$1645$^b$ &--	&$-$0.66$\pm$0.11 &$-$1.14$\pm$0.09 &$-$1.54$\pm$0.11 &$-$1.34$\pm$0.08 &$-$0.96$\pm$0.08 &$<-$2.11$^a$     &$-$0.36$\pm$0.14&--\\
Q2335$+$1501 &$>-$2.55	&--		  &$-$0.44$\pm$0.40 &--		      &$-$0.32$\pm$0.33	&$+$0.07$\pm$0.34 &$<-$1.23	    &$-$1.60$\pm$0.33&$-$0.04$\pm$0.15\\
\hline\hline 				       			 	 
\end{tabular}			       			 	 
\end{center}			       			 	 
\begin{minipage}{140mm}
{\bf $^a$:} 3-$\sigma$ limit calculated from the non-detection of the line.\\
{\bf $^b$:} Q0449$-$1645 was not observed by SDSS, therefore no $\Delta$(g-i) could be calculated.\\
\end{minipage}
\end{table*}

\subsection{On the Velocity-spread/Metallicity Relation}

In three of the six absorbers studied here, we could determine the equivalent
width of \mgii\ \l\ 2796 from the high-resolution UVES data. For these systems we can thus derive the D-indices defined by Ellison (2006), where, as in the original work,  $\Delta v$ is the total line width excluding 'detached' absorption components where the continuum is recovered and only including the complex profile and the EW is expressed in \AA:

\begin{equation}
D=1000 \times EW (\mgii\ 2796) /  \Delta v
\end{equation}
 
This gives $D$-index=5.9, 5.6 and 5.4 for absorbers towards Q0138$-$0005, Q0153$+$0009 and Q2335$+$1501 respectively. So these three sub-DLAs, indeed have a $D$-index below the 6.3 limit indicating \lognhi$<$20.3.
 
On the other hand, it has been recently advocated that the well-known galaxy luminosity-metallicity relation observed at $0<z<1$ (Tremonti et al. 2004)  was already in place at higher-redshifts. In fact, a mass-metallicity relation has recently been put into evidence for UV-selected star-forming galaxies at $z\sim 2.3$ by Erb et al. (2006). An empirical relation between velocity spread and metallicity in quasar absorbers was independently reported by P\'eroux et al. (2003b). More recently, Ledoux et al. (2006), have used this relation to claim a mass-metallicity relation in high-redshift absorbers, {\it assuming} that the velocity width of quasar absorbers is a good proxy for halo masses.  Meiring et al. (2007) also compared the metallicity-velocity width correlation for sub-DLAs and DLAs, as did Prochaska et al. (2007) for the GRB host galaxies. 

We note that this empirical correlation is expected to relate to the evolution of the $D$-index with \lognhi.  However, the interpretation of $\Delta v$ as a measure of the halo mass (Haehnelt, Steinmetz \& Rauch 1998; Maller et al.2001) is challenged by observational evidences. In fact, recent \hi-gas rich observations of local galaxies by Zwaan et al. (submitted),  show that the velocity spread is not a good indicator of mass in \hi-gas rich DLA analogues at z=0. Similarly, Bouch\'e et al. (2007) have recently suggested that the equivalent width of \mgii\ absorbers is due to winds, not gravity related velocity dispersion.

\section{Conclusions}

We have used new high-resolution observations of six quasar absorbers to constrain the metallicity of sub-DLAs in the low-redshift Universe. We specially targeted the \znii\ line which is expected to be a metallicity indicator free from the bias effect of dust. In line with recent results from our group (Meiring \e\ 2007, 2008; P\'eroux \e\ 2006a, 2006b; Khare \e\ 2007; Kulkarni \e\ 2007), we find that sub-DLAs appear to be more metal-rich than classical DLAs. In particular, we report the discovery of two sub-DLAs with super-solar metallicity towards Q0138$-$0005 with [Zn/H]=$+$0.28$\pm$0.16 and Q2335$+$1501 with [Zn/H]=$+$0.07$\pm$0.34. We note that there are significant variations in column density ratios of metals from component to component in some of our systems, and also that there is some indication of different grain types being required if the depletion is related to extinction. While it is still possible that  the abundance variation depends on uncertain ionisation corrections, current photo-ionisation models (Meiring  \e\ 2007, 2008) do suggest the variations are in abundance, not radiation fields being different from component to component.

\section*{Acknowledgements}
We thank Sandhya Rao for providing \nhi\ estimate of the absorber towards Q2335$+$1501 in advance of publication. We would like to thank the Paranal and Garching staff at ESO for
performing the observations in Service Mode. VPK and JM acknowledge partial support from the U.S. National Science Foundation grant AST-0607739 (PI: Kulkarni).

\bsp

\label{lastpage}

\end{document}